\documentclass[a4paper,UKenglish,cleveref,autoref,thm-restate]{lipics-v2021}
\usepackage{algorithm,algpseudocode,amsthm,refcount}
\newcommand{\BWT}{\ensuremath{\mathrm{BWT}}}
\newcommand{\SA}{\ensuremath{\mathrm{SA}}}
\newcommand{\LCP}{\ensuremath{\mathrm{LCP}}}
\newcommand{\context}{\ensuremath{\mathrm{context}}}

\title{Merging RLBWTs adaptively}

\author{Travis Gagie}{Faculty of Computer Science, Dalhousie University, Canada}{travis.gagie@gmail.com}{https://orcid.org/0000-0003-3689-327X}{Funded by NSERC Discovery Grant RGPIN-07185-2020.}

\authorrunning{T. Gagie}

\Copyright{Travis Gagie}



\nolinenumbers

\EventEditors{Philip Bille and Nicola Prezza}
\EventNoEds{2}
\EventLongTitle{37th Annual Symposium on Combinatorial Pattern Matching (CPM 2026)}
\EventShortTitle{CPM 2026}
\EventAcronym{CPM}
\EventYear{2026}
\EventDate{June 15--17, 2026}
\EventLocation{Copenhagen, Denmark}
\EventLogo{}
\SeriesVolume{369}
\ArticleNo{20}

\begin{document}

\maketitle

\begin{abstract}
We show how to merge two run-length compressed Burrows-Wheeler Transforms (RLBWTs) into a run-length compressed extended Burrows-Wheeler Transform (eBWT) in $O (r)$ space and $O ((r + L) \log (m + n))$ time, where $m$ and $n$ are the lengths of the uncompressed strings, $r$ is the number of runs in the final eBWT and $L$ is the sum of its irreducible LCP values. 
\end{abstract}

\section{Introduction}
\label{sec:introduction}

The Burrows-Wheeler Transform (BWT)~\cite{BW94} is an important tool in modern genomics.  As DNA sequencing technologies have advanced and researchers have started working with pangenomic references, they have begun using run-length compressed BWTs (RLBWTs).  There are good algorithms for building RLBWTs for massive pangenomic references in which all the genomes are similar~\cite{BGKLMM19,Kem19} but the case when the reference contains many genomes from each of many species is still challenging.  One possible solution is to build a separate RLBWT for each species and then merge them into a single RLBWT or run-length compressed extended BWT (eBWT)~\cite{MRRS07}.

Merging BWTs is an established research topic~\cite{FGM12,HM14,Sir16} and merging RLBWTs~\cite{ORSMKGB21,Li24} is a natural extension of that.  As far as we know, however, until recently all algorithms for merging RLBWTs were based on dynamic RLBWTs~\cite{OSTIS18,BGI20} and used time at least roughly linear in the size of the uncompressed inputs, even if they used compressed space.  D\'iaz-Dom\'inguez et al.~\cite{DGGLLMMS25} gave an algorithm for merging BWTs that can be extended to RLBWTs and works faster when the inputs are individually repetitive but not similar to each other.  Because it is based on prefix-free parsing, however, it does not have good worst-case bounds.  In this paper we show how to merge two RLBWTs into a run-length compressed eBWT in $O (r)$ space and $O ((r + L) \log (m + n))$ time, where $m$ and $n$ are the lengths of the uncompressed strings, $r$ is the number of runs in the final eBWT and $L$ is the sum of the longest common prefix (LCP) values at the beginnings of those runs (known as its irreducible LCP values).  It is known that $L \in O ((m + n) \log \delta)$~\cite{KK22}, where $\delta \leq r$ is a powerful measure of the eBWT's compressibility~\cite{KNP22}.

We describe here only how to merge the RLBWTs of two strings.  Our algorithm can easily be extended to merge many RLBWTs or run-length compressed eBWTs, but we leave that to the full version of the paper.  In Section~\ref{sec:preliminaries} we review some preliminary concepts.  As warm-ups, in Section~\ref{sec:warm-ups} we present simple algorithms for merging positional BWTs (PBWTs) and RLBWTs.  In Section~\ref{sec:algorithm} we present our main algorithm and in Section~\ref{sec:analysis} we present a first analysis that yields a time bound of $O ((r \log r + L) \log (m + n))$.  In Section~\ref{sec:construction} we give the proof of a technical lemma we use, whose details we refer to in Section~\ref{sec:optimization} when we optimize our algorithm slightly to reduce the $r \log r$ in our time bound to $r$, so the whole bound becomes $O ((r + L) \log (m + n))$.

\section{Preliminaries}
\label{sec:preliminaries}

For the sake of brevity, we assume readers are familiar with run-length compression, the Burrows-Wheeler Transform (BWT), run-length compressed BWTs (RLBWTs), positional BWTs (PBWTs)~\cite{Dur14}, suffix arrays (SAs), the $\Psi$ function and longest common prefix (LCP) arrays; see M\"akinen et al.'s~\cite{MBCT23} and Navarro's~\cite{Nav16} texts for an introduction.  The extended BWT (eBWT)~\cite{MRRS07} of two strings $S [1..m]$ and $T [1..n]$ is an interleaving $\BWT_{S, T} [1..m + n]$ of the characters $\BWT_S [1..m]$ and $\BWT_T [1..n]$ of $S$ and $T$.  Since $\BWT_S$ and $\BWT_T$ are subsequences of $\BWT_{S, T}$, if it has $r$ runs then they each have at most $r$ runs.  For convenience, we write $\BWT_S$, $\BWT_T$ and $\BWT_{S, T}$ to denote both the BWTs and the RLBWTs, stating when they are run-length compressed.

For $1 \leq h \leq m + n$, $\BWT_{S, T} [h]$ has the lexicographically $h$th smallest context.  The contexts of $\BWT_S [i]$ and $\BWT_T [j]$ are
\begin{eqnarray*}
\context_S (i) & = & S [\SA_S [i]..m] \circ S [1..\SA_S [i] - 1] \\
\context_T (j) & = & T [\SA_T [j]..n] \circ T [1..\SA_T [j] - 1]
\end{eqnarray*}
(unless $\SA_S [i] = 1$, in which case $\context_S (i)$ is just $S [1..m]$, or $\SA_T [j] = 1$, in which case $\context_T (j)$ is just $T [1..n]$), where $\circ$ denotes concatenation.  If we assume $S$ and $T$ are each terminated by an end-of-string symbol that occurs nowhere else in $S$ and $T$ then two characters in the same BWT cannot have the same context, and a character in $\BWT_S$ and a character in $\BWT_T$ have the same context if and only if $S = T$ and the characters are in the same positions (in which case we can break ties arbitrarily without affecting $\BWT_{S, T}$).

We can store a move structure~\cite{NT21} (see also~\cite{BGR22}) for the $\Psi$ function $\Psi_S$ of $S$ in $O (r)$ space such that, given $\BWT_S$ and position $i$ in it, we can extract $\BWT_S [i]$'s context
\[S [\SA_S [i]..n] \circ S [1..\SA_S [i] - 1]
= \BWT_S [\Psi_S (i)]\,\BWT_S [\Psi_S^2 (i)]\,\BWT_S [\Psi_S^3 (i)] \cdots \BWT_S [\Psi_S^n (i)]\]
character by character by iteratively evaluating $\Psi_S$ in $O (\log r)$ time plus constant time per iteration or, equivalently, per character extracted.  Brown~\cite{Bro23} and Bertram et al.~\cite{BFN24} (see also~\cite{BL26}) showed how we can build such a move structure in $O (r \log r)$ time.  An analogous result holds for $T$.

\begin{lemma}
\label{lem:psi}
Given the RLBWTs $\BWT_S$ and $\BWT_T$, in $O (r \log r)$ time we can build $O (r)$-space data structures for iteratively evaluating the $\Psi$ functions for $S$ and $T$ in $O (\log r)$ time plus constant time per iteration.
\end{lemma}

\noindent It follows that we can compare $\context_S (i)$ and $\context_T (j)$ in $O (\log r)$ time plus time proportional to the length of their longest common prefix.  We give a proof of Lemma~\ref{lem:psi} in Section~\ref{sec:construction} and then explain in Section~\ref{sec:optimization} how we can pay the $O (\log r)$ overhead only three times during a binary search to find, for example, the smallest value $j$ with $\context_T (j)$ lexicographically larger than $\context_S (i)$, denoted $\context_T (j) \succ \context_S (i)$, rather than paying it at every step of the search.

\section{Warm-ups}
\label{sec:warm-ups}

Suppose we have the PBWTs for two sets of haplotypes from the same species, with the same number of columns (but not necessarily the same number of rows) and with the columns representing the same variation sites, and we want the PBWT for all the haplotypes.  For example, consider the PBWTs shown in Figure~\ref{fig:PBWTs}, with the bottom PBWT being the merge of the top two (and the example from~\cite{ICGVCGBB25} with indexing from 1 for consistency with the rest of this paper).  The entries of the prefix arrays (showing what haplotype each PBWT entry corresponds to) are shown in parentheses.  Cells are red if their information is for one of the first 10 haplotypes and blue if it is for one of the second 10.  This makes it easy to see that,  by the definition of the PBWT, each column in the merged PBWT is an interleaving of the corresponding columns in the two input PBWTs.

\begin{figure}[t]
\resizebox{\textwidth}{!}
{\begin{tabular}{*{14}{c@{\hspace{1ex}}r@{\hspace{4ex}}}c@{\hspace{1ex}}r}
1 && 2 && 3 && 4 && 5 && 6 && 7 && 8 && 9 && 10 && 11 && 12 && 13 && 14 && 15 \\
\hline
 \color{red} 1 &  \color{red}  (1) &  \color{red} 1 &  \color{red}  (5) &  \color{red} 0 &  \color{red}  (1) &  \color{red} 1 &  \color{red}  (1) &  \color{red} 1 &  \color{red}  (9) &  \color{red} 0 &  \color{red}  (1) &  \color{red} 0 &  \color{red}  (1) &  \color{red} 0 &  \color{red}  (1) &  \color{red} 0 &  \color{red}  (1) &  \color{red} 0 &  \color{red}  (1) &  \color{red} 0 &  \color{red}  (1) &  \color{red} 1 &  \color{red}  (1) &  \color{red} 1 &  \color{red}  (8) &  \color{red} 1 &  \color{red}  (2) &  \color{red} 1 &  \color{red}  (5) \\
 \color{red} 1 &  \color{red}  (2) &  \color{red} 1 &  \color{red}  (6) &  \color{red} 0 &  \color{red}  (2) &  \color{red} 1 &  \color{red}  (2) &  \color{red} 0 &  \color{red}  (1) &  \color{red} 1 &  \color{red}  (5) &  \color{red} 0 &  \color{red} (10) &  \color{red} 0 &  \color{red} (10) &  \color{red} 1 &  \color{red} (10) &  \color{red} 1 &  \color{red}  (9) &  \color{red} 0 &  \color{red}  (5) &  \color{red} 1 &  \color{red}  (5) &  \color{red} 0 &  \color{red}  (2) &  \color{red} 1 &  \color{red} (10) &  \color{red} 1 &  \color{red}  (6) \\
 \color{red} 1 &  \color{red}  (3) &  \color{red} 1 &  \color{red}  (7) &  \color{red} 0 &  \color{red}  (3) &  \color{red} 1 &  \color{red}  (3) &  \color{red} 1 &  \color{red}  (2) &  \color{red} 1 &  \color{red}  (6) &  \color{red} 0 &  \color{red}  (9) &  \color{red} 0 &  \color{red}  (9) &  \color{red} 0 &  \color{red}  (9) &  \color{red} 0 &  \color{red}  (5) &  \color{red} 0 &  \color{red}  (6) &  \color{red} 1 &  \color{red}  (6) &  \color{red} 0 &  \color{red} (10) &  \color{red} 0 &  \color{red}  (5) &  \color{red} 1 &  \color{red}  (7) \\
 \color{red} 1 &  \color{red}  (4) &  \color{red} 1 &  \color{red}  (8) &  \color{red} 0 &  \color{red}  (4) &  \color{red} 1 &  \color{red}  (4) &  \color{red} 1 &  \color{red}  (3) &  \color{red} 1 &  \color{red}  (7) &  \color{red} 0 &  \color{red}  (2) &  \color{red} 1 &  \color{red}  (2) &  \color{red} 0 &  \color{red}  (5) &  \color{red} 0 &  \color{red}  (6) &  \color{red} 0 &  \color{red}  (7) &  \color{red} 1 &  \color{red}  (7) &  \color{red} 1 &  \color{red}  (1) &  \color{red} 0 &  \color{red}  (6) &  \color{red} 1 &  \color{red}  (3) \\
 \color{red} 0 &  \color{red}  (5) &  \color{red} 1 &  \color{red}  (9) &  \color{red} 0 &  \color{red}  (5) &  \color{red} 1 &  \color{red}  (5) &  \color{red} 1 &  \color{red}  (4) &  \color{red} 1 &  \color{red}  (8) &  \color{red} 0 &  \color{red}  (3) &  \color{red} 1 &  \color{red}  (3) &  \color{red} 0 &  \color{red}  (6) &  \color{red} 0 &  \color{red}  (7) &  \color{red} 0 &  \color{red}  (8) &  \color{red} 0 &  \color{red}  (8) &  \color{red} 0 &  \color{red}  (5) &  \color{red} 0 &  \color{red}  (7) &  \color{red} 1 &  \color{red}  (4) \\
 \color{red} 0 &  \color{red}  (6) &  \color{red} 1 &  \color{red} (10) &  \color{red} 0 &  \color{red}  (6) &  \color{red} 1 &  \color{red}  (6) &  \color{red} 0 &  \color{red}  (5) &  \color{red} 0 &  \color{red} (10) &  \color{red} 0 &  \color{red}  (4) &  \color{red} 1 &  \color{red}  (4) &  \color{red} 0 &  \color{red}  (7) &  \color{red} 0 &  \color{red}  (8) &  \color{red} 0 &  \color{red}  (2) &  \color{red} 0 &  \color{red}  (2) &  \color{red} 0 &  \color{red}  (6) &  \color{red} 0 &  \color{red}  (3) &  \color{red} 1 &  \color{red}  (9) \\
 \color{red} 0 &  \color{red}  (7) &  \color{red} 0 &  \color{red}  (1) &  \color{red} 0 &  \color{red}  (7) &  \color{red} 1 &  \color{red}  (7) &  \color{red} 0 &  \color{red}  (6) &  \color{red} 0 &  \color{red}  (9) &  \color{red} 0 &  \color{red}  (5) &  \color{red} 0 &  \color{red}  (5) &  \color{red} 0 &  \color{red}  (8) &  \color{red} 0 &  \color{red}  (2) &  \color{red} 0 &  \color{red}  (3) &  \color{red} 1 &  \color{red}  (3) &  \color{red} 0 &  \color{red}  (7) &  \color{red} 0 &  \color{red}  (4) &  \color{red} 1 &  \color{red}  (8) \\
 \color{red} 0 &  \color{red}  (8) &  \color{red} 0 &  \color{red}  (2) &  \color{red} 0 &  \color{red}  (8) &  \color{red} 1 &  \color{red}  (8) &  \color{red} 0 &  \color{red}  (7) &  \color{red} 0 &  \color{red}  (2) &  \color{red} 0 &  \color{red}  (6) &  \color{red} 0 &  \color{red}  (6) &  \color{red} 0 &  \color{red}  (2) &  \color{red} 0 &  \color{red}  (3) &  \color{red} 0 &  \color{red}  (4) &  \color{red} 1 &  \color{red}  (4) &  \color{red} 0 &  \color{red}  (3) &  \color{red} 0 &  \color{red}  (9) &  \color{red} 1 &  \color{red}  (1) \\
 \color{red} 0 &  \color{red}  (9) &  \color{red} 0 &  \color{red}  (3) &  \color{red} 0 &  \color{red}  (9) &  \color{red} 0 &  \color{red}  (9) &  \color{red} 0 &  \color{red}  (8) &  \color{red} 0 &  \color{red}  (3) &  \color{red} 0 &  \color{red}  (7) &  \color{red} 0 &  \color{red}  (7) &  \color{red} 0 &  \color{red}  (3) &  \color{red} 0 &  \color{red}  (4) &  \color{red} 0 &  \color{red} (10) &  \color{red} 0 &  \color{red} (10) &  \color{red} 0 &  \color{red}  (4) &  \color{red} 0 &  \color{red}  (8) &  \color{red} 1 &  \color{red}  (2) \\
 \color{red} 0 &  \color{red} (10) &  \color{red} 0 &  \color{red}  (4) &  \color{red} 0 &  \color{red} (10) &  \color{red} 1 &  \color{red} (10) &  \color{red} 0 &  \color{red} (10) &  \color{red} 0 &  \color{red}  (4) &  \color{red} 0 &  \color{red}  (8) &  \color{red} 0 &  \color{red}  (8) &  \color{red} 0 &  \color{red}  (4) &  \color{red} 0 &  \color{red} (10) &  \color{red} 1 &  \color{red}  (9) &  \color{red} 1 &  \color{red}  (9) &  \color{red} 0 &  \color{red}  (9) &  \color{red} 0 &  \color{red}  (1) &  \color{red} 1 &  \color{red} (10) \\
\end{tabular}}

\bigskip

\resizebox{\textwidth}{!}
{\begin{tabular}{*{14}{c@{\hspace{1ex}}r@{\hspace{4ex}}}c@{\hspace{1ex}}r}
1 && 2 && 3 && 4 && 5 && 6 && 7 && 8 && 9 && 10 && 11 && 12 && 13 && 14 && 15 \\
\hline
\color{blue} 0 & \color{blue} (11) & \color{blue} 1 & \color{blue} (11) & \color{blue} 0 & \color{blue} (11) & \color{blue} 1 & \color{blue} (11) & \color{blue} 1 & \color{blue} (12) & \color{blue} 0 & \color{blue} (15) & \color{blue} 0 & \color{blue} (15) & \color{blue} 0 & \color{blue} (15) & \color{blue} 1 & \color{blue} (15) & \color{blue} 0 & \color{blue} (17) & \color{blue} 0 & \color{blue} (17) & \color{blue} 1 & \color{blue} (17) & \color{blue} 1 & \color{blue} (20) & \color{blue} 1 & \color{blue} (11) & \color{blue} 1 & \color{blue} (19) \\
\color{blue} 0 & \color{blue} (12) & \color{blue} 1 & \color{blue} (12) & \color{blue} 0 & \color{blue} (12) & \color{blue} 0 & \color{blue} (12) & \color{blue} 1 & \color{blue} (13) & \color{blue} 0 & \color{blue} (16) & \color{blue} 0 & \color{blue} (16) & \color{blue} 0 & \color{blue} (16) & \color{blue} 1 & \color{blue} (16) & \color{blue} 0 & \color{blue} (12) & \color{blue} 1 & \color{blue} (12) & \color{blue} 1 & \color{blue} (19) & \color{blue} 1 & \color{blue} (15) & \color{blue} 0 & \color{blue} (19) & \color{blue} 0 & \color{blue} (12) \\
\color{blue} 0 & \color{blue} (13) & \color{blue} 1 & \color{blue} (13) & \color{blue} 0 & \color{blue} (13) & \color{blue} 0 & \color{blue} (13) & \color{blue} 1 & \color{blue} (14) & \color{blue} 1 & \color{blue} (18) & \color{blue} 0 & \color{blue} (11) & \color{blue} 0 & \color{blue} (11) & \color{blue} 1 & \color{blue} (11) & \color{blue} 0 & \color{blue} (19) & \color{blue} 0 & \color{blue} (19) & \color{blue} 1 & \color{blue} (18) & \color{blue} 1 & \color{blue} (16) & \color{blue} 0 & \color{blue} (12) & \color{blue} 1 & \color{blue} (13) \\
\color{blue} 0 & \color{blue} (14) & \color{blue} 1 & \color{blue} (14) & \color{blue} 0 & \color{blue} (14) & \color{blue} 0 & \color{blue} (14) & \color{blue} 0 & \color{blue} (15) & \color{blue} 0 & \color{blue} (11) & \color{blue} 0 & \color{blue} (17) & \color{blue} 0 & \color{blue} (17) & \color{blue} 0 & \color{blue} (17) & \color{blue} 0 & \color{blue} (18) & \color{blue} 0 & \color{blue} (18) & \color{blue} 0 & \color{blue} (20) & \color{blue} 0 & \color{blue} (11) & \color{blue} 0 & \color{blue} (13) & \color{blue} 1 & \color{blue} (14) \\
\color{blue} 0 & \color{blue} (15) & \color{blue} 1 & \color{blue} (15) & \color{blue} 0 & \color{blue} (15) & \color{blue} 0 & \color{blue} (15) & \color{blue} 0 & \color{blue} (16) & \color{blue} 0 & \color{blue} (17) & \color{blue} 0 & \color{blue} (12) & \color{blue} 0 & \color{blue} (12) & \color{blue} 0 & \color{blue} (12) & \color{blue} 0 & \color{blue} (20) & \color{blue} 0 & \color{blue} (20) & \color{blue} 0 & \color{blue} (15) & \color{blue} 1 & \color{blue} (17) & \color{blue} 0 & \color{blue} (14) & \color{blue} 1 & \color{blue} (20) \\
\color{blue} 0 & \color{blue} (16) & \color{blue} 1 & \color{blue} (16) & \color{blue} 0 & \color{blue} (16) & \color{blue} 0 & \color{blue} (16) & \color{blue} 0 & \color{blue} (18) & \color{blue} 0 & \color{blue} (12) & \color{blue} 0 & \color{blue} (13) & \color{blue} 0 & \color{blue} (13) & \color{blue} 1 & \color{blue} (13) & \color{blue} 0 & \color{blue} (15) & \color{blue} 0 & \color{blue} (15) & \color{blue} 0 & \color{blue} (16) & \color{blue} 0 & \color{blue} (19) & \color{blue} 0 & \color{blue} (20) & \color{blue} 1 & \color{blue} (15) \\
\color{blue} 0 & \color{blue} (17) & \color{blue} 1 & \color{blue} (17) & \color{blue} 0 & \color{blue} (17) & \color{blue} 1 & \color{blue} (17) & \color{blue} 1 & \color{blue} (19) & \color{blue} 0 & \color{blue} (13) & \color{blue} 0 & \color{blue} (14) & \color{blue} 0 & \color{blue} (14) & \color{blue} 1 & \color{blue} (14) & \color{blue} 0 & \color{blue} (16) & \color{blue} 0 & \color{blue} (16) & \color{blue} 0 & \color{blue} (11) & \color{blue} 1 & \color{blue} (18) & \color{blue} 0 & \color{blue} (15) & \color{blue} 1 & \color{blue} (16) \\
\color{blue} 1 & \color{blue} (18) & \color{blue} 1 & \color{blue} (19) & \color{blue} 1 & \color{blue} (19) & \color{blue} 0 & \color{blue} (18) & \color{blue} 1 & \color{blue} (20) & \color{blue} 0 & \color{blue} (14) & \color{blue} 0 & \color{blue} (19) & \color{blue} 0 & \color{blue} (19) & \color{blue} 0 & \color{blue} (19) & \color{blue} 0 & \color{blue} (11) & \color{blue} 0 & \color{blue} (11) & \color{blue} 1 & \color{blue} (12) & \color{blue} 0 & \color{blue} (12) & \color{blue} 0 & \color{blue} (16) & \color{blue} 1 & \color{blue} (17) \\
\color{blue} 0 & \color{blue} (19) & \color{blue} 1 & \color{blue} (20) & \color{blue} 1 & \color{blue} (20) & \color{blue} 0 & \color{blue} (19) & \color{blue} 0 & \color{blue} (11) & \color{blue} 0 & \color{blue} (19) & \color{blue} 1 & \color{blue} (20) & \color{blue} 0 & \color{blue} (18) & \color{blue} 0 & \color{blue} (18) & \color{blue} 0 & \color{blue} (13) & \color{blue} 1 & \color{blue} (13) & \color{blue} 1 & \color{blue} (13) & \color{blue} 0 & \color{blue} (13) & \color{blue} 0 & \color{blue} (17) & \color{blue} 1 & \color{blue} (18) \\
\color{blue} 0 & \color{blue} (20) & \color{blue} 1 & \color{blue} (18) & \color{blue} 0 & \color{blue} (18) & \color{blue} 0 & \color{blue} (20) & \color{blue} 0 & \color{blue} (17) & \color{blue} 0 & \color{blue} (20) & \color{blue} 0 & \color{blue} (18) & \color{blue} 0 & \color{blue} (20) & \color{blue} 0 & \color{blue} (20) & \color{blue} 0 & \color{blue} (14) & \color{blue} 1 & \color{blue} (14) & \color{blue} 1 & \color{blue} (14) & \color{blue} 0 & \color{blue} (14) & \color{blue} 0 & \color{blue} (18) & \color{blue} 1 & \color{blue} (11) \\
\end{tabular}}

\bigskip

\resizebox{\textwidth}{!}
{\begin{tabular}{*{14}{c@{\hspace{1ex}}r@{\hspace{4ex}}}c@{\hspace{1ex}}r}
1 && 2 && 3 && 4 && 5 && 6 && 7 && 8 && 9 && 10 && 11 && 12 && 13 && 14 && 15 \\
\hline
 \color{red} 1 &  \color{red}  (1) &  \color{red} 1 &  \color{red}  (5) &  \color{red} 0 &  \color{red}  (1) &  \color{red} 1 &  \color{red}  (1) &  \color{red} 1 &  \color{red}  (9) & \color{blue} 0 & \color{blue} (15) & \color{blue} 0 & \color{blue} (15) & \color{blue} 0 & \color{blue} (15) & \color{blue} 1 & \color{blue} (15) &  \color{red} 0 &  \color{red}  (1) &  \color{red} 0 &  \color{red}  (1) &  \color{red} 1 &  \color{red}  (1) &  \color{red} 1 &  \color{red}  (8) &  \color{red} 1 &  \color{red}  (2) & \color{blue} 1 & \color{blue} (19) \\
 \color{red} 1 &  \color{red}  (2) &  \color{red} 1 &  \color{red}  (6) &  \color{red} 0 &  \color{red}  (2) &  \color{red} 1 &  \color{red}  (2) & \color{blue} 1 & \color{blue} (12) & \color{blue} 0 & \color{blue} (16) & \color{blue} 0 & \color{blue} (16) & \color{blue} 0 & \color{blue} (16) & \color{blue} 1 & \color{blue} (16) & \color{blue} 0 & \color{blue} (17) & \color{blue} 0 & \color{blue} (17) & \color{blue} 1 & \color{blue} (17) & \color{blue} 1 & \color{blue} (20) &  \color{red} 1 &  \color{red} (10) &  \color{red} 1 &  \color{red}  (5) \\
 \color{red} 1 &  \color{red}  (3) &  \color{red} 1 &  \color{red}  (7) &  \color{red} 0 &  \color{red}  (3) &  \color{red} 1 &  \color{red}  (3) & \color{blue} 1 & \color{blue} (13) & \color{blue} 1 & \color{blue} (18) &  \color{red} 0 &  \color{red}  (1) &  \color{red} 0 &  \color{red}  (1) &  \color{red} 0 &  \color{red}  (1) &  \color{red} 1 &  \color{red}  (9) & \color{blue} 1 & \color{blue} (12) & \color{blue} 1 & \color{blue} (19) &  \color{red} 0 &  \color{red}  (2) & \color{blue} 1 & \color{blue} (11) &  \color{red} 1 &  \color{red}  (6) \\
 \color{red} 1 &  \color{red}  (4) &  \color{red} 1 &  \color{red}  (8) &  \color{red} 0 &  \color{red}  (4) &  \color{red} 1 &  \color{red}  (4) & \color{blue} 1 & \color{blue} (14) &  \color{red} 0 &  \color{red}  (1) &  \color{red} 0 &  \color{red} (10) &  \color{red} 0 &  \color{red} (10) &  \color{red} 1 &  \color{red} (10) & \color{blue} 0 & \color{blue} (12) & \color{blue} 0 & \color{blue} (19) & \color{blue} 1 & \color{blue} (18) & \color{blue} 1 & \color{blue} (15) & \color{blue} 0 & \color{blue} (19) &  \color{red} 1 &  \color{red}  (7) \\
 \color{red} 0 &  \color{red}  (5) &  \color{red} 1 &  \color{red}  (9) &  \color{red} 0 &  \color{red}  (5) &  \color{red} 1 &  \color{red}  (5) & \color{blue} 0 & \color{blue} (15) &  \color{red} 1 &  \color{red}  (5) & \color{blue} 0 & \color{blue} (11) & \color{blue} 0 & \color{blue} (11) & \color{blue} 1 & \color{blue} (11) & \color{blue} 0 & \color{blue} (19) & \color{blue} 0 & \color{blue} (18) &  \color{red} 1 &  \color{red}  (5) & \color{blue} 1 & \color{blue} (16) &  \color{red} 0 &  \color{red}  (5) &  \color{red} 1 &  \color{red}  (3) \\
 \color{red} 0 &  \color{red}  (6) &  \color{red} 1 &  \color{red} (10) &  \color{red} 0 &  \color{red}  (6) &  \color{red} 1 &  \color{red}  (6) & \color{blue} 0 & \color{blue} (16) &  \color{red} 1 &  \color{red}  (6) & \color{blue} 0 & \color{blue} (17) & \color{blue} 0 & \color{blue} (17) & \color{blue} 0 & \color{blue} (17) & \color{blue} 0 & \color{blue} (18) &  \color{red} 0 &  \color{red}  (5) &  \color{red} 1 &  \color{red}  (6) &  \color{red} 0 &  \color{red} (10) &  \color{red} 0 &  \color{red}  (6) &  \color{red} 1 &  \color{red}  (4) \\
 \color{red} 0 &  \color{red}  (7) & \color{blue} 1 & \color{blue} (11) &  \color{red} 0 &  \color{red}  (7) &  \color{red} 1 &  \color{red}  (7) & \color{blue} 0 & \color{blue} (18) &  \color{red} 1 &  \color{red}  (7) &  \color{red} 0 &  \color{red}  (9) &  \color{red} 0 &  \color{red}  (9) &  \color{red} 0 &  \color{red}  (9) &  \color{red} 0 &  \color{red}  (5) &  \color{red} 0 &  \color{red}  (6) &  \color{red} 1 &  \color{red}  (7) & \color{blue} 0 & \color{blue} (11) &  \color{red} 0 &  \color{red}  (7) & \color{blue} 0 & \color{blue} (12) \\
 \color{red} 0 &  \color{red}  (8) & \color{blue} 1 & \color{blue} (12) &  \color{red} 0 &  \color{red}  (8) &  \color{red} 1 &  \color{red}  (8) & \color{blue} 1 & \color{blue} (19) &  \color{red} 1 &  \color{red}  (8) & \color{blue} 0 & \color{blue} (12) & \color{blue} 0 & \color{blue} (12) & \color{blue} 0 & \color{blue} (12) &  \color{red} 0 &  \color{red}  (6) &  \color{red} 0 &  \color{red}  (7) &  \color{red} 0 &  \color{red}  (8) &  \color{red} 1 &  \color{red}  (1) &  \color{red} 0 &  \color{red}  (3) & \color{blue} 1 & \color{blue} (13) \\
 \color{red} 0 &  \color{red}  (9) & \color{blue} 1 & \color{blue} (13) &  \color{red} 0 &  \color{red}  (9) &  \color{red} 0 &  \color{red}  (9) & \color{blue} 1 & \color{blue} (20) &  \color{red} 0 &  \color{red} (10) & \color{blue} 0 & \color{blue} (13) & \color{blue} 0 & \color{blue} (13) & \color{blue} 1 & \color{blue} (13) &  \color{red} 0 &  \color{red}  (7) &  \color{red} 0 &  \color{red}  (8) & \color{blue} 0 & \color{blue} (20) & \color{blue} 1 & \color{blue} (17) &  \color{red} 0 &  \color{red}  (4) & \color{blue} 1 & \color{blue} (14) \\
 \color{red} 0 &  \color{red} (10) & \color{blue} 1 & \color{blue} (14) &  \color{red} 0 &  \color{red} (10) &  \color{red} 1 &  \color{red} (10) &  \color{red} 0 &  \color{red}  (1) & \color{blue} 0 & \color{blue} (11) & \color{blue} 0 & \color{blue} (14) & \color{blue} 0 & \color{blue} (14) & \color{blue} 1 & \color{blue} (14) &  \color{red} 0 &  \color{red}  (8) & \color{blue} 0 & \color{blue} (20) &  \color{red} 0 &  \color{red}  (2) & \color{blue} 0 & \color{blue} (19) & \color{blue} 0 & \color{blue} (12) &  \color{red} 1 &  \color{red}  (9) \\
\color{blue} 0 & \color{blue} (11) & \color{blue} 1 & \color{blue} (15) & \color{blue} 0 & \color{blue} (11) & \color{blue} 1 & \color{blue} (11) &  \color{red} 1 &  \color{red}  (2) & \color{blue} 0 & \color{blue} (17) & \color{blue} 0 & \color{blue} (19) & \color{blue} 0 & \color{blue} (19) & \color{blue} 0 & \color{blue} (19) & \color{blue} 0 & \color{blue} (20) &  \color{red} 0 &  \color{red}  (2) &  \color{red} 1 &  \color{red}  (3) & \color{blue} 1 & \color{blue} (18) & \color{blue} 0 & \color{blue} (13) &  \color{red} 1 &  \color{red}  (8) \\
\color{blue} 0 & \color{blue} (12) & \color{blue} 1 & \color{blue} (16) & \color{blue} 0 & \color{blue} (12) & \color{blue} 0 & \color{blue} (12) &  \color{red} 1 &  \color{red}  (3) &  \color{red} 0 &  \color{red}  (9) & \color{blue} 1 & \color{blue} (20) &  \color{red} 1 &  \color{red}  (2) & \color{blue} 0 & \color{blue} (18) &  \color{red} 0 &  \color{red}  (2) &  \color{red} 0 &  \color{red}  (3) &  \color{red} 1 &  \color{red}  (4) &  \color{red} 0 &  \color{red}  (5) & \color{blue} 0 & \color{blue} (14) & \color{blue} 1 & \color{blue} (20) \\
\color{blue} 0 & \color{blue} (13) & \color{blue} 1 & \color{blue} (17) & \color{blue} 0 & \color{blue} (13) & \color{blue} 0 & \color{blue} (13) &  \color{red} 1 &  \color{red}  (4) & \color{blue} 0 & \color{blue} (12) &  \color{red} 0 &  \color{red}  (2) &  \color{red} 1 &  \color{red}  (3) &  \color{red} 0 &  \color{red}  (5) &  \color{red} 0 &  \color{red}  (3) &  \color{red} 0 &  \color{red}  (4) & \color{blue} 0 & \color{blue} (15) &  \color{red} 0 &  \color{red}  (6) &  \color{red} 0 &  \color{red}  (9) & \color{blue} 1 & \color{blue} (15) \\
\color{blue} 0 & \color{blue} (14) & \color{blue} 1 & \color{blue} (19) & \color{blue} 0 & \color{blue} (14) & \color{blue} 0 & \color{blue} (14) &  \color{red} 0 &  \color{red}  (5) & \color{blue} 0 & \color{blue} (13) &  \color{red} 0 &  \color{red}  (3) &  \color{red} 1 &  \color{red}  (4) &  \color{red} 0 &  \color{red}  (6) &  \color{red} 0 &  \color{red}  (4) & \color{blue} 0 & \color{blue} (15) & \color{blue} 0 & \color{blue} (16) &  \color{red} 0 &  \color{red}  (7) &  \color{red} 0 &  \color{red}  (8) & \color{blue} 1 & \color{blue} (16) \\
\color{blue} 0 & \color{blue} (15) & \color{blue} 1 & \color{blue} (20) & \color{blue} 0 & \color{blue} (15) & \color{blue} 0 & \color{blue} (15) &  \color{red} 0 &  \color{red}  (6) & \color{blue} 0 & \color{blue} (14) &  \color{red} 0 &  \color{red}  (4) & \color{blue} 0 & \color{blue} (18) &  \color{red} 0 &  \color{red}  (7) & \color{blue} 0 & \color{blue} (15) & \color{blue} 0 & \color{blue} (16) &  \color{red} 0 &  \color{red} (10) &  \color{red} 0 &  \color{red}  (3) & \color{blue} 0 & \color{blue} (20) &  \color{red} 1 &  \color{red}  (1) \\
\color{blue} 0 & \color{blue} (16) &  \color{red} 0 &  \color{red}  (1) & \color{blue} 0 & \color{blue} (16) & \color{blue} 0 & \color{blue} (16) &  \color{red} 0 &  \color{red}  (7) & \color{blue} 0 & \color{blue} (19) & \color{blue} 0 & \color{blue} (18) &  \color{red} 0 &  \color{red}  (5) &  \color{red} 0 &  \color{red}  (8) & \color{blue} 0 & \color{blue} (16) &  \color{red} 0 &  \color{red} (10) & \color{blue} 0 & \color{blue} (11) &  \color{red} 0 &  \color{red}  (4) & \color{blue} 0 & \color{blue} (15) & \color{blue} 1 & \color{blue} (17) \\
\color{blue} 0 & \color{blue} (17) &  \color{red} 0 &  \color{red}  (2) & \color{blue} 0 & \color{blue} (17) & \color{blue} 1 & \color{blue} (17) &  \color{red} 0 &  \color{red}  (8) & \color{blue} 0 & \color{blue} (20) &  \color{red} 0 &  \color{red}  (5) &  \color{red} 0 &  \color{red}  (6) & \color{blue} 0 & \color{blue} (20) &  \color{red} 0 &  \color{red} (10) & \color{blue} 0 & \color{blue} (11) & \color{blue} 1 & \color{blue} (12) & \color{blue} 0 & \color{blue} (12) & \color{blue} 0 & \color{blue} (16) & \color{blue} 1 & \color{blue} (18) \\
\color{blue} 1 & \color{blue} (18) &  \color{red} 0 &  \color{red}  (3) & \color{blue} 1 & \color{blue} (19) & \color{blue} 0 & \color{blue} (18) &  \color{red} 0 &  \color{red} (10) &  \color{red} 0 &  \color{red}  (2) &  \color{red} 0 &  \color{red}  (6) &  \color{red} 0 &  \color{red}  (7) &  \color{red} 0 &  \color{red}  (2) & \color{blue} 0 & \color{blue} (11) & \color{blue} 1 & \color{blue} (13) & \color{blue} 1 & \color{blue} (13) & \color{blue} 0 & \color{blue} (13) &  \color{red} 0 &  \color{red}  (1) &  \color{red} 1 &  \color{red}  (2) \\
\color{blue} 0 & \color{blue} (19) &  \color{red} 0 &  \color{red}  (4) & \color{blue} 1 & \color{blue} (20) & \color{blue} 0 & \color{blue} (19) & \color{blue} 0 & \color{blue} (11) &  \color{red} 0 &  \color{red}  (3) &  \color{red} 0 &  \color{red}  (7) &  \color{red} 0 &  \color{red}  (8) &  \color{red} 0 &  \color{red}  (3) & \color{blue} 0 & \color{blue} (13) & \color{blue} 1 & \color{blue} (14) & \color{blue} 1 & \color{blue} (14) & \color{blue} 0 & \color{blue} (14) & \color{blue} 0 & \color{blue} (17) &  \color{red} 1 &  \color{red} (10) \\
\color{blue} 0 & \color{blue} (20) & \color{blue} 1 & \color{blue} (18) & \color{blue} 0 & \color{blue} (18) & \color{blue} 0 & \color{blue} (20) & \color{blue} 0 & \color{blue} (17) &  \color{red} 0 &  \color{red}  (4) &  \color{red} 0 &  \color{red}  (8) & \color{blue} 0 & \color{blue} (20) &  \color{red} 0 &  \color{red}  (4) & \color{blue} 0 & \color{blue} (14) &  \color{red} 1 &  \color{red}  (9) &  \color{red} 1 &  \color{red}  (9) &  \color{red} 0 &  \color{red}  (9) & \color{blue} 0 & \color{blue} (18) & \color{blue} 1 & \color{blue} (11) \\
\end{tabular}}
\caption{Two PBWTs for 10 haplotypes each {\bf (above)}, and the PBWT for all 20 haplotypes {\bf (below)}.  The entries of the prefix arrays are shown in parentheses.  Cells are red if their information is for one of the first 10 haplotypes and blue if it is for one of the second 10.}
\label{fig:PBWTs}
\end{figure}

The first column from the left of the merged PBWT is the concatenation of the the first columns of the two input PBWTs.  Suppose we have already merged the $(j - 1)$st columns of the input PBWTs into the $(j - 1)$st column of the merged PBWT, and now we want to merge their $j$th columns into its $j$th column.  In particular, suppose we know how the $(j - 1)$st column of the merged PBWT is divided into maximal blocks of consecutive bits that all come from the same input PBWT.  In our example, the blocks in the 9th column have lengths $2, 2, 2, 1, 5, 4, 1, 3$ and alternate between blue and red with the first block being blue.

To compute the $j$th column of the merged PBWT after computing the $(j - 1)$st column, we first consider all the 0s in the $(j - 1)$st column and list the next bits in their haplotypes (in the 0s' order in the $(j - 1)$st column), then consider all the 1s in the $(j - 1)$st column and list the next bits in their haplotypes (in the 1s' order in the $(j - 1)$st column); the $j$th column of the merged PBWT is the concatenation of the two lists.  By induction, the bits in the $j$th column are in the co-lexicographic order of the preceding prefixes in their haplotypes.

Because the $j$th columns of the two input PBWTs already contain all those next bits and in the correct order, we can use the input PBWTs instead of the explicit haplotypes.  To do this, we start with an empty draft of the $j$th column of the merged PBWT and scan the $(j - 1)$st column of the merged PBWT twice.  During the first scan, whenever we see a 0 we copy a bit (the next bit in that 0's haplotype) from the $j$th column of the input PBWT that is the source of that 0.  During the second scan, whenever we see a 1 we copy a bit (the next bit in that 1's haplotype) from the $j$th column of the input PBWT that is the source of that 1.  However, our algorithm may work more efficiently considering blocks rather than literally bit by bit.

Suppose the $(j - 1)$st column of the merged PBWT consists of $b$ blocks.  For $k$ from 1 to $b$, if there are $c$ copies of 0 in the $k$th block in the $(j - 1)$st column then we append to our draft of the $j$th column of the merged PBWT the next $c$ bits of the $j$th column of the input PBWT that is the source of the bits in the $k$th block.  After that, for $k$ from 1 to $b$, if there are $c$ copies of 1 in the $k$th block of the $(j - 1)$st column then we append to our draft of the $j$th column of the merged PBWT the next $c$ bits of the $j$th column of the input PBWT that is the source of the bits in the $k$th block.

In our example, there are no copies of 0s in the first block of the 9th column, which is blue, so we append no bits from the 10th column of the blue PBWT; there is 1 copy of 0 in the second block of the 9th column, which is red, so we append 1 bit (\textcolor{red}{0}) from the 10th column of the red PBWT (from haplotype \textcolor{red}{(1)}); then 1 bit (\textcolor{blue}{0}) from the blue PBWT (from haplotype \textcolor{blue}{(17)}); then 1 bit (\textcolor{red}{1}) from the red PBWT (from haplotype \textcolor{red}{(9)}); then 3 bits (\textcolor{blue}{000}) from the blue PBWT (from haplotypes \textcolor{blue}{(12)}, \textcolor{blue}{(19)} and \textcolor{blue}{(18)}); then 4 bits (\textcolor{red}{0000}) from the red PBWT (from haplotypes \textcolor{red}{(5)}, \textcolor{red}{(6)}, \textcolor{red}{(7)} and \textcolor{red}{(8)}); then 1 bit (\textcolor{blue}{0}) from the blue PBWT (from haplotype \textcolor{blue}{(20)}); then 3 bits (\textcolor{red}{000}) from the red PBWT (from haplotypes \textcolor{red}{(2)}, \textcolor{red}{(3)} and \textcolor{red}{(4)}).

There are 2 copies of 1 in the first block of the 9th column, which is blue, so we append 2 bits (\textcolor{blue}{00}) from the 10th column of the blue PBWT (from haplotypes \textcolor{blue}{(15)} and \textcolor{blue}{(16)}); then 1 bit (\textcolor{red}{0}) from the from the red PBWT (from haplotype \textcolor{red}{(10)}); then 1 bit (\textcolor{blue}{0}) from the blue PBWT (from haplotype \textcolor{blue}{(11)}); then no bits from the red PBWT; then 2 bits (\textcolor{blue}{00}) from the blue PBWT (from haplotypes \textcolor{blue}{(13)} and \textcolor{blue}{(14)}).  When we are finished, we know the 10th column of the merged PBWT is \textcolor{red}{0}\textcolor{blue}{0}\textcolor{red}{1}\textcolor{blue}{000}\textcolor{red}{0000}\textcolor{blue}{0}\textcolor{red}{000}\textcolor{blue}{00}\textcolor{red}{0}\textcolor{blue}{000} and we know its blocks have lengths 1, 1, 3, 4, 1, 3, 2, 1, 3; its prefix array is \textcolor{red}{(1)} \textcolor{blue}{(17)} \textcolor{red}{(9)} \textcolor{blue}{(12) (19) (18)} \textcolor{red}{(5) (6) (7) (8)} \textcolor{blue}{(20)} \textcolor{red}{(2) (3) (4)} \textcolor{blue}{(15) (16)} \textcolor{red}{(10)} \textcolor{blue}{(11) (13) (14)}.

If a block in the $(j - 1)$st column of the merged PBWT overlaps $t$ runs of bits in that column, then computing the numbers of 0s and 1s in that block takes $O (t)$ time.  We perform one append operation for the 0s in that block and another for the 1s.  If the input PBWTs are run-length compressed and we want only the $j$th column and its block structure --- which is all we need to continue merging the two input PBWTs --- but not its prefix array, then each append operation takes time proportional to the number of runs in the substring of bits we append.  It follows that we can merge two PBWTs in time $O (r + B)$, where $r$ and $B$ are the total numbers of runs and blocks in all the columns of the merged PBWT, respectively.

\begin{theorem}
\label{thm:PBWTs}
We can merge two PBWTs with run-length compressed columns in $O (r + B)$ time, where $r$ and $B$ are the total numbers of runs and blocks in all the columns of the merged PBWT, respectively.
\end{theorem}

\noindent
Theorem~\ref{thm:PBWTs} may be of independent interest but we have proven it mainly as a warm-up.  As a second warm-up, we now give a simple algorithm for merging RLBWTs that uses $O (r)$ space and $O \left( \rule{0ex}{2ex} r \log r + (b \log r + B + \sigma) \log (m + n) \right)$ time, where $r$ is the number of runs in the resulting eBWT, $b$ is the number of maximal blocks of characters in that eBWT that all come from the same input RLBWT, $B$ is the sum of the LCP values at the beginnings of those blocks, $\sigma$ is the size of the alphabet, and $m$ and $n$ are the lengths of the uncompressed input BWTs.

Suppose we are to merge the RLBWTs $\BWT_S [1..m]$ and $\BWT_T [1..n]$ of strings $S [1..m]$ and $T [1..n]$ into the run-length compressed eBWT $\BWT_{S, T} [1..m + n]$ of $S$ and $T$ together, and we have already built the data structures for $\Psi$ for $S$ and $T$ described in Lemma~\ref{lem:psi} so that we can compare characters' contexts.  For example, if $S$ is the concatenation of the \$-terminated strings GATTACAT, GATACAT and GATTAGATA and $T$ is the concatenation of those three strings' \$-terminated reverse complements, then
\begin{eqnarray*}
\BWT_S & = & \mathrm{TAT^5C^2G^4A^3\$^3A^3TATA^2} \\
\BWT_T & = & \mathrm{C^3T^2A^2T^2\$^2T^6CG^2\$A^6} \\
\BWT_{S, T} & = & \mathrm{C^2TATCT^6C^2G^2A^2T^2\$^2G^2T^3A^2TA\$^3T^2A^3CGTATG\$A^8}\,,
\end{eqnarray*}
with exponents indicating run lengths, and $\BWT_S$ and $\BWT_T$ are interleaved in $\BWT_{S, T}$ as shown in Figure~\ref{fig:RLBWTs}.  In this toy example, $r = 27$, $b = 18$, $B = 18$, $\sigma = 5$ and $m + n = 54$.

\begin{figure}[t]
\resizebox{\textwidth}{!}
{\begin{tabular}{cc}
\begin{tabular}{rl@{\hspace{.5ex}}c}
0 & \textcolor{blue}{\$ATGTAATC\$ATGTATC\$TATCTAAT} & \textcolor{blue}{C} \\
  & \textcolor{blue}{\$ATGTATC\$TATCTAATC\$ATGTAAT} & \textcolor{blue}{C} \\
1 & \textcolor{red}{\$GATACAT\$GATTAGATA\$GATTACA} & \textcolor{red}{T} \\
  & \textcolor{red}{\$GATTACAT\$GATACAT\$GATTAGAT} & \textcolor{red}{A} \\
  & \textcolor{red}{\$GATTAGATA\$GATTACAT\$GATACA} & \textcolor{red}{T} \\
1 & \textcolor{blue}{\$TATCTAATC\$ATGTAATC\$ATGTAT} & \textcolor{blue}{C} \\
0 & \textcolor{red}{A\$GATTACAT\$GATACAT\$GATTAGA} & \textcolor{red}{T} \\
1 & \textcolor{blue}{AATC\$ATGTAATC\$ATGTATC\$TATC} & \textcolor{blue}{T} \\
  & \textcolor{blue}{AATC\$ATGTATC\$TATCTAATC\$ATG} & \textcolor{blue}{T} \\
1 & \textcolor{red}{ACAT\$GATACAT\$GATTAGATA\$GAT} & \textcolor{red}{T} \\
  & \textcolor{red}{ACAT\$GATTAGATA\$GATTACAT\$GA} & \textcolor{red}{T} \\
  & \textcolor{red}{AGATA\$GATTACAT\$GATACAT\$GAT} & \textcolor{red}{T} \\
  & \textcolor{red}{AT\$GATACAT\$GATTAGATA\$GATTA} & \textcolor{red}{C} \\
  & \textcolor{red}{AT\$GATTAGATA\$GATTACAT\$GATA} & \textcolor{red}{C} \\
  & \textcolor{red}{ATA\$GATTACAT\$GATACAT\$GATTA} & \textcolor{red}{G} \\
  & \textcolor{red}{ATACAT\$GATTAGATA\$GATTACAT\$} & \textcolor{red}{G} \\
2 & \textcolor{blue}{ATC\$ATGTAATC\$ATGTATC\$TATCT} & \textcolor{blue}{A} \\
  & \textcolor{blue}{ATC\$ATGTATC\$TATCTAATC\$ATGT} & \textcolor{blue}{A} \\
  & \textcolor{blue}{ATC\$TATCTAATC\$ATGTAATC\$ATG} & \textcolor{blue}{T} \\
  & \textcolor{blue}{ATCTAATC\$ATGTAATC\$ATGTATC\$} & \textcolor{blue}{T} \\
  & \textcolor{blue}{ATGTAATC\$ATGTATC\$TATCTAATC} & \textcolor{blue}{\$} \\
  & \textcolor{blue}{ATGTATC\$TATCTAATC\$ATGTAATC} & \textcolor{blue}{\$} \\
2 & \textcolor{red}{ATTACAT\$GATACAT\$GATTAGATA\$} & \textcolor{red}{G} \\
  & \textcolor{red}{ATTAGATA\$GATTACAT\$GATACAT\$} & \textcolor{red}{G} \\
0 & \textcolor{blue}{C\$ATGTAATC\$ATGTATC\$TATCTAA} & \textcolor{blue}{T} \\
  & \textcolor{blue}{C\$ATGTATC\$TATCTAATC\$ATGTAA} & \textcolor{blue}{T} \\
  & \textcolor{blue}{C\$TATCTAATC\$ATGTAATC\$ATGTA} & \textcolor{blue}{T}
\end{tabular}
&
\begin{tabular}{rl@{\hspace{.5ex}}c}
1 & \textcolor{red}{CAT\$GATACAT\$GATTAGATA\$GATT} & \textcolor{red}{A} \\
  & \textcolor{red}{CAT\$GATTAGATA\$GATTACAT\$GAT} & \textcolor{red}{A} \\
1 & \textcolor{blue}{CTAATC\$ATGTAATC\$ATGTATC\$TA} & \textcolor{blue}{T} \\
0 & \textcolor{red}{GATA\$GATTACAT\$GATACAT\$GATT} & \textcolor{red}{A} \\
  & \textcolor{red}{GATACAT\$GATTAGATA\$GATTACAT} & \textcolor{red}{\$} \\
  & \textcolor{red}{GATTACAT\$GATACAT\$GATTAGATA} & \textcolor{red}{\$} \\
  & \textcolor{red}{GATTAGATA\$GATTACAT\$GATACAT} & \textcolor{red}{\$} \\
1 & \textcolor{blue}{GTAATC\$ATGTATC\$TATCTAATC\$A} & \textcolor{blue}{T} \\
  & \textcolor{blue}{GTATC\$TATCTAATC\$ATGTAATC\$A} & \textcolor{blue}{T} \\
0 & \textcolor{red}{T\$GATACAT\$GATTAGATA\$GATTAC} & \textcolor{red}{A} \\
  & \textcolor{red}{T\$GATTAGATA\$GATTACAT\$GATAC} & \textcolor{red}{A} \\
  & \textcolor{red}{TA\$GATTACAT\$GATACAT\$GATTAG} & \textcolor{red}{A} \\
2 & \textcolor{blue}{TAATC\$ATGTAATC\$ATGTATC\$TAT} & \textcolor{blue}{C} \\
  & \textcolor{blue}{TAATC\$ATGTATC\$TATCTAATC\$AT} & \textcolor{blue}{G} \\
2 & \textcolor{red}{TACAT\$GATACAT\$GATTAGATA\$GA} & \textcolor{red}{T} \\
  & \textcolor{red}{TACAT\$GATTAGATA\$GATTACAT\$G} & \textcolor{red}{A} \\
  & \textcolor{red}{TAGATA\$GATTACAT\$GATACAT\$GA} & \textcolor{red}{T} \\
2 & \textcolor{blue}{TATC\$TATCTAATC\$ATGTAATC\$AT} & \textcolor{blue}{G} \\
  & \textcolor{blue}{TATCTAATC\$ATGTAATC\$ATGTATC} & \textcolor{blue}{\$} \\
  & \textcolor{blue}{TC\$ATGTAATC\$ATGTATC\$TATCTA} & \textcolor{blue}{A} \\
  & \textcolor{blue}{TC\$ATGTATC\$TATCTAATC\$ATGTA} & \textcolor{blue}{A} \\
  & \textcolor{blue}{TC\$TATCTAATC\$ATGTAATC\$ATGT} & \textcolor{blue}{A} \\
  & \textcolor{blue}{TCTAATC\$ATGTAATC\$ATGTATC\$T} & \textcolor{blue}{A} \\
  & \textcolor{blue}{TGTAATC\$ATGTATC\$TATCTAATC\$} & \textcolor{blue}{A} \\
  & \textcolor{blue}{TGTATC\$TATCTAATC\$ATGTAATC\$} & \textcolor{blue}{A} \\
1 & \textcolor{red}{TTACAT\$GATACAT\$GATTAGATA\$G} & \textcolor{red}{A} \\
  & \textcolor{red}{TTAGATA\$GATTACAT\$GATACAT\$G} & \textcolor{red}{A}
\end{tabular}
\end{tabular}}
\caption{The lexicographically sorted cyclic shifts of the concatenation of the three \$-terminated strings GATTACAT, GATACAT and GATTAGATA {\bf (in red)}, and of the concatenation of those three strings' \$-terminated reverse complements {\bf (in blue)}, with the LCP values shown at block boundaries {\bf (in black)}.  The eBWT of the two concatenations is the last column of the matrix, slightly set from the rest.  The red subsequence of the eBWT is the BWT of the first concatenation and the blue subsequence is the BWT of the second.}
\label{fig:RLBWTs}
\end{figure}

We first set $i$ and $j$ to 1, then compare $\context_S (i)$ and $\context_T (j)$ to check whether the first block of characters in $\BWT_{S, T}$ is from $\BWT_S$ or $\BWT_T$, and set a Boolean flag to true in the former case and false in the latter.  As long as $i \leq m$ and $j \leq n$, if the flag is true then we use a doubling search in $\BWT_S [i..m]$ to find the largest value $i'$ with $\context_S (i') \prec \context_T (j)$, output $\BWT_S [i..i']$, reset $i$ to $i' + 1$, and reset the flag to false.  If the flag is false then we use a doubling search in $\BWT_T [j..n]$ to find the largest value $j'$ with $\context_T (j') \prec \context_S (i)$, output $\BWT_T [j..j']$, reset $j$ to $j' + 1$, and reset the flag to true.  Eventually, when $i = m + 1$ or $j = n + 1$, we output either $\BWT_T [j..n]$ or $\BWT_S [i..m]$, respectively.  The pseudocode for our algorithm is shown in Figure~\ref{fig:warm-up_pseudocode}.

\begin{figure}[t]
\begin{algorithmic}[1]
\State $i \gets 1$
\State $j \gets 1$
\If{$\context_S (i) \prec \context_T (j)$}
    \State $\mathit{flag} \gets \mathrm{true}$
\Else
    \State $\mathit{flag} \gets \mathrm{false}$
\EndIf
\While{$i \leq m$ and $j \leq n$}
	\If{$\mathit{flag}$}
        \State doubling search for the largest value $i'$ with $\context_S (i') \prec \context_T (j)$
		\State output $\BWT_S [i..i']$
		\State $i \gets i' + 1$
        \State $\mathit{flag} \gets \mathrm{false}$
	\Else
        \State doubling search for the largest value $j'$ with $\context_T (j') \prec \context_S (i)$
		\State output $\BWT_T [j..j']$
        \State $j \gets j' + 1$
        \State $\mathit{flag} \gets \mathrm{true}$
	\EndIf
\EndWhile
\If{$\mathit{flag}$}
	\State output $\BWT_S [i..m]$
\Else
	\State output $\BWT_T [j..n]$
\EndIf
\end{algorithmic}
\caption{Pseudocode for merging $\BWT_S [1..m]$ and $\BWT_T [1..n]$ into $\BWT_{S, T} [1..m + n]$.}
\label{fig:warm-up_pseudocode}
\end{figure}

By induction, during each pass through the while loop, we output the next block of $\BWT_{S, T}$, from $\BWT_S$ if the flag is true and from $\BWT_T$ if the flag is false.  If the flag is true then all of the $O (\log m)$ comparisons we perform during the current pass are against $\context_T (j)$, which is in the block from $\BWT_T$ we will output during the next pass, so the number of characters we extract for each of those comparisons is at most 1 plus the maximum of the LCP values at the boundaries of that block from $\BWT_T$.  If the flag is false then all of the $O (\log n)$ comparisons we perform during the current pass are against $\context_S (i)$, which is in the block from $\BWT_S$ we will output during the next pass, so the number of characters we extract for each of those comparisons is at most 1 plus the maximum of the LCP values at the boundaries of the block from $\BWT_S$.

It follows that we can charge all the comparisons to blocks of $\BWT_{S, T}$ such that each block is charged $O (\log (m + n))$ comparisons and for each comparison charged to a block, the number of characters we extract is at most 1 plus the maximum of the LCP values at the boundaries of that block.  Since at most $\sigma$ of those LCP values are 0, this means we use $O \left( \rule{0ex}{2ex} r \log r + (b \log r + B + \sigma) \log (m + n) \right)$ total time, including the $O (r \log r)$ time to build the data structures for comparing contexts and the $O (\log r)$ overhead for each of the $O (b \log (m + n))$ comparisons.  Our algorithm is adaptive in the sense that it takes advantage of cases when $b$ and $B$ are small, which is likely when $S$ and $T$ are individually repetitive but dissimilar.

\begin{theorem}
\label{thm:simple}
Given the RLBWTs $\BWT_S [1..m]$ and $\BWT_T [1..n]$, we can merge them into the run-length compressed eBWT $\BWT_{S, T} [1..m + n]$ using $O (r)$ space and $O \left( \rule{0ex}{2ex}\!\right.\!r \log r + (b \log r + B + \sigma) \log (m + n)\!\left.\!\rule{0ex}{2ex} \right)$ time, where $r$ is the number of runs in the resulting eBWT, $b$ is the number of maximal blocks of characters in that eBWT that all come from the same input RLBWT, $B$ is the sum of the LCP values at the beginnings of those blocks, $\sigma$ is the size of the alphabet, and $m$ and $n$ are the lengths of the uncompressed input BWTs.
\end{theorem}

\section{Main algorithm}
\label{sec:algorithm}

Again, suppose we are to merge the RLBWTs $\BWT_S [1..m]$ and $\BWT_T [1..n]$ of strings $S [1..m]$ and $T [1..n]$ into the run-length compressed eBWT $\BWT_{S, T} [1..m + n]$ of $S$ and $T$ together, and we have already built the data structures for $\Psi$ for $S$ and $T$ described in Lemma~\ref{lem:psi} so that we can compare characters' contexts.  The pseudocode for our algorithm is shown in Figure~\ref{fig:main_pseudocode}.

\begin{figure}[p]
\resizebox{.98\textwidth}{!}
{\begin{minipage}{\textwidth}
\begin{algorithmic}[1]
\State $i \gets 1$
\State $j \gets 1$
\While{$i \leq m$ and $j \leq n$}
	\State $k \gets \mbox{length of leading run of $\BWT_S [i..m]$}$
	\State $\ell \gets \mbox{length of leading run of $\BWT_T [j..n]$}$
	\If{$\BWT_S [i] = \BWT_T [j]$}
		\If{$i + k \leq m$ and $\context_S (i + k) \prec \context_T (j + \ell - 1)$}
			\State binary search for the smallest value $j'$ with $\context_T (j') \succ \context_S (i + k)$
			\State output $\BWT_S [i..i + k - 1] \circ \BWT_T [j..j' - 1]$
			\State $i \gets i + k$
			\State $j \gets j'$
		\ElsIf{$j + \ell \leq n$ and $\context_T (j + \ell) \prec \context_S (i + k - 1)$}
			\State binary search for the smallest value $i'$ with $\context_S (i') \succ \context_T (j + \ell)$
			\State output $\BWT_S [i..i' - 1] \circ \BWT_T [j..j + \ell - 1]$
			\State $i \gets i'$
			\State $j \gets j + \ell$
		\Else
			\State output $\BWT_S [i..i + k - 1] \circ \BWT_T [j..j + \ell - 1]$
			\State $i \gets i + k$
			\State $j \gets j + \ell$	
		\EndIf
	\Else
		\If{$\context_S (i) \prec \context_T (j)$}
			\If{$\context_S (i + k - 1) \prec \context_T (j)$}
				\State output $\BWT_S [i..i + k - 1]$
				\State $i \gets i + k$
			\Else
				\State binary search for the smallest value $i'$ with $\context_S (i') \succ \context_T (j)$
				\State output $\BWT_S [i..i' - 1]$
				\State $i \gets i'$
			\EndIf
		\Else
			\If{$\context_T (j + \ell - 1) \prec \context_S (i)$}				
				\State output $\BWT_T [j..j + \ell - 1]$
				\State $j \gets j + \ell$
			\Else
				\State binary search for the smallest value $j'$ with $\context_T (j') \succ \context_S (i)$
				\State output $\BWT_T [j..j' - 1]$
				\State $j \gets j'$
			\EndIf
		\EndIf
	\EndIf
\EndWhile
\If{$i \leq m$}
	\State output $\BWT_S [i..m]$
\ElsIf{$j \leq n$}
	\State output $\BWT_T [j..n]$
\EndIf
\end{algorithmic}
\end{minipage}}
\caption{Pseudocode for merging $\BWT_S [1..m]$ and $\BWT_T [1..n]$ into $\BWT_{S, T} [1..m + n]$.}
\label{fig:main_pseudocode}
\end{figure}

Assume we have already merged $\BWT_S [1..i - 1]$ and $\BWT_T [1..j - 1]$ into $\BWT_{S, T} [1..i + j - 2]$ correctly.  When we first reach the while loop (lines 3--43 in Figure~\ref{fig:main_pseudocode}) $i = j = 1$ so this assumption is trivially true.  If $i > m$ then $\BWT_{S, T} [i + j - 1..m + n] = \BWT_T [j..n]$ (line 45) and if $j > n$ then $\BWT_{S, T} [i + j - 1..m + n] = \BWT_S [i..m]$ (line 47), so assume $i \leq m$ and $j \leq n$ and consider only the while loop.  Let $k$ and $\ell$ be the lengths of the leading runs of $\BWT_S [i..m]$ and $\BWT_T [j..n]$ respectively (lines 4 and 5).

First suppose $\BWT_S [i] = \BWT_T [j]$ (lines 7--21), meaning the leading runs $\BWT_S [i..i + k - 1]$ and $\BWT_T [j..j + \ell - 1]$ of $\BWT_S [i..m]$ and $\BWT_T [j..n]$ consist of copies of the same character.  If there exists a character $\BWT_S [i + k]$ after the run $\BWT_S [i..i + k - 1]$ and $\context_S (i + k) \prec \context_T (j + \ell - 1)$ (lines 8--11) then $\BWT_S [i + k]$ precedes some non-empty suffix $\BWT_T [j'..j + \ell - 1]$ of the run $\BWT_T [j..j + \ell - 1]$ in $\BWT_{S, T}$, with $j'$ being the smallest value with $\context_T (j') \succ \context_S (i + k)$.  We use binary search to find $j'$ in the range $[j, j + \ell)$ (line 8), output
\[\BWT_S [i..i + k - 1] \circ \BWT_T [j..j' - 1]
= (\BWT_S [i])^{k + j' - j}\]
(line 9) and reset $i$ to $i + k$ (line 10) and $j$ to $j'$ (line 11).  Since the characters we output are all equal, we need not consider their order to merge the RLBWTs.  (We note, however, that we would need to consider the characters' order if we were also merging the RLBWTs' SAs.)  The case when there exists a character $\BWT_T [j + \ell]$ after the run $\BWT_T [j..j + \ell - 1]$ and $\context_T (j + \ell) \prec \context_S (i + k - 1)$ (lines 13--16) is symmetric.

By the definition of the eBWT, it is impossible for $\BWT_S [i + k]$ to precede a non-empty suffix of the run $\BWT_T [j..j + \ell - 1]$ in $\BWT_{S, T}$ and for $\BWT_T [j + \ell]$ to precede a non-empty suffix of the run $\BWT_S [i..i + k - 1]$ in $\BWT_{S, T}$ in $\BWT_{S, T}$, since then $\BWT_S [i + k]$ would have to both precede and follow $\BWT_T [j + \ell]$.  Therefore, the only remaining case with $\BWT_S [i] = \BWT_T [j]$ is when either $i + k = m + 1$ or $\BWT_S [i + k]$ follows the entire run $\BWT_T [j..j + \ell - 1]$ in $\BWT_{S, T}$ and either $j + \ell = n + 1$ or $\BWT_T [j + \ell]$ follows the entire run $\BWT_S [i..i + k - 1]$ in $\BWT_{S, T}$.  In this case, we output
\[\BWT_S [i..i + k - 1] \circ \BWT_T [j..j + \ell - 1]
= (\BWT_S [i])^{k + \ell}\]
(line 18) and reset $i$ to $i + k$ (line 19) and $j$ to $j + \ell$ (line 20). 

Now suppose $\BWT_S [i] \neq \BWT_T [j]$ (lines 23--42).  Because $S [\SA_S [i] - 1] = \BWT_S [i]$ (or $S [m] = \BWT_S [i]$ if $\SA_S [i] = 1$ so $S [1..\SA_S [i] - 1]$ is empty) and $T [\SA_T [j] - 1] = \BWT_T [j]$ (or $T [n] = \BWT_T [j]$ if $\SA_T [j] = 1$ so $T [1..\SA_T [j] - 1]$ is empty), $\context_S (i)$ and $\context_T (j)$ differ on their last characters.  Therefore, we consider only two cases: when $\context_S (i) \prec \context_T (j)$ (lines 24--31) and when $\context_S (i) \succ \context_T (j)$ (lines 33--40).

Suppose $\context_S (i) \prec \context_T (j)$, meaning $\BWT_S [i]$ precedes $\BWT_T [j]$ in $\BWT_{S, T}$.  If $\context_S (i + k - 1) \prec \context_T (j)$ then the entire run $\BWT_S [i..i + k - 1]$ precedes $\BWT_T [j]$ in $\BWT_{S, T}$, so we output that run (line 25) and reset $i$ to $i + k$ (line 26).  Otherwise, some non-empty prefix $\BWT_S [i..i' - 1]$ of the run $\BWT_S [i..i + k - 1]$ precedes $\BWT_T [j]$ in $\BWT_{S, T}$ and some non-empty suffix $\BWT_S [i'..i + k - 1]$ follows it, with $i'$ being the smallest value with $\context_S (i') \succ \context_T (j)$.  We use binary search (line 28) to find $i'$ in the range $(i, i + k - 1]$, comparing the context of a character in $\BWT_S [i + 1..i + k - 1]$ to $\context_T (j)$ at each step.  We output $\BWT_S [i..i' - 1]$ (line 29) and reset $i$ to $i'$ (line 30).  The case when $\context_S (i) \succ \context_T (j)$ is symmetric.

We exit the while loop only after having consumed all of at least one of $\BWT_S$ and $\BWT_T$.  If we have not consumed all of the other, we output the rest of it (lines 44 to 48).

\section{First analysis}
\label{sec:analysis}

Building the structures in Lemma~\ref{lem:psi} takes $O (r \log r) \subseteq O (r \log (m + n))$ time, where $r$ is the number of runs in $\BWT_{S, T}$.  Since $\BWT_S$ and $\BWT_T$ both consist of at most $r$ runs, outputting $\BWT_S [i..m]$ (line 45) or $\BWT_T [j..n]$ (line 47) in run-length compressed form takes $O (r)$ time.  Therefore, we can focus on the complexity of the while loop.

\begin{lemma}
\label{lem:runs}
During each pass through the while loop, we output an entire run in $\BWT_{S, T}$.
\end{lemma}

\begin{proof}
Whenever we reach line 9 we have $\context_S (i + k) \prec \context_T (j')$ so, after we output
\[\BWT_S [i..i + k - 1] \circ \BWT_T [j..j' - 1]
= (\BWT_S [i])^{k + j' - j}\,,\]
the next character we output is $\BWT_S [i + k] \neq \BWT_S [i]$ and it starts a new run in $\BWT_{S, T}$.  Symmetrically, after we reach line 14 and output
\[\BWT_S [i..i' - 1] \circ \BWT_T [j..j + \ell - 1]
= (\BWT_S [i])^{i' - i + \ell}\,,\]
the next character we output is $\BWT_T [j + \ell] \neq \BWT_T [j]$ and it starts a new run in $\BWT_{S, T}$.  After we reach line 18 and output
\[\BWT_S [i..i + k - 1] \circ \BWT_T [j..j + \ell - 1]
= (\BWT_S [i])^{k + \ell}\,,\]
the next character we output (if there is one) is either $\BWT_S [i + k] \neq \BWT_S [i]$ or $\BWT_T [j + \ell] \neq \BWT_T [j]$ and it starts a new run in $\BWT_{S, T}$.  After we reach line 25 and output
\[\BWT_S [i..i + k - 1]
= (\BWT_S [i])^k\,,\]
the next character we output is either $\BWT_S [i + k] \neq \BWT_S [i]$ or $\BWT_T [j] \neq \BWT_S [i]$ and it starts a new run in $\BWT_{S, T}$.  Symmetrically, after we reach line 34 and output
\[\BWT_T [j..j + \ell - 1]
= (\BWT_T [j])^\ell\,,\]
the next character we output is either $\BWT_T [i + \ell] \neq \BWT_T [j]$ or $\BWT_S [i] \neq \BWT_T [j]$ and it starts a new run in $\BWT_{S, T}$.  After we reach line 29 and output
\[\BWT_S [i..i' - 1]
= (\BWT_S [i])^{i' - i}\,,\]
the next character we output is $\BWT_T [j] \neq \BWT_S [i]$ and it starts a new run in $\BWT_{S, T}$.  Symmetrically, after we reach line 38 and output
\[\BWT_T [j..j' - 1]
= (\BWT_T [j])^{j' - j}\,,\]
the next character we output is $\BWT_S [i] \neq \BWT_T [j]$ and it starts a new run in $\BWT_{S, T}$.
\end{proof}

\noindent It follows from Lemma~\ref{lem:runs} that we make $O (r)$ passes through the while loop.  Since $\BWT_S$ and $\BWT_T$ and $\BWT_{S, T}$ are run-length compressed, the only operations that can take more than constant time during a single pass through the while loop are comparisons between characters' contexts, of which we make $O (\log (m + n))$ during each pass, so $O (r \log (m + n))$ in total.

\begin{lemma}
\label{lem:comparisons}
Whenever we compare two characters' contexts
\begin{itemize}
\item those two characters are not equal, so they are in different runs in $\BWT_{S, T}$;
\item at least one of those two characters is in the run in $\BWT_{S, T}$ output during the current pass through the while loop or the run in $\BWT_{S, T}$ output during the next pass.
\end{itemize}
\end{lemma}

\begin{proof}
When we check whether $\context_S (i + k) \prec \context_T (j + \ell - 1)$ in line 7,
\[\BWT_S [i + k] \neq \BWT_S [i] = \BWT_T [j] = \BWT_T [j + \ell - 1]\]
and we output either $\BWT_S [i + k]$ or $\BWT_T [j + \ell]$ during the next pass.  The comparison in line 12 is symmetric.  When we perform the binary search in line 8, in each step we check whether $\context_T (j') \succ \context_S (i + k)$ for some $j' < j + \ell$ with
\[\BWT_T [j'] = \BWT_T [j] = \BWT_S [i] \neq \BWT_S [i + k]\]
and we output $\BWT_S [i + k]$ during the next pass.  The binary search in line 13 is symmetric.  When we check whether $\context_S (i) \prec \context_T (j)$ in line 23, $\BWT_S [i] \neq \BWT_T [j]$ and we output either $\BWT_S [i]$ or $\BWT_T [j]$ during the current pass.  When we check whether $\context_S (i + k - 1) \prec \context_T (j)$ in line 24,
\[\BWT_S [i + k - 1] = \BWT_S [i] \neq \BWT_T [j]\]
and we output either $\BWT_S [i + k - 1]$ during the current pass or $\BWT_T [j]$ during the next pass.  The comparison in line 33 is symmetric.  When we perform the binary search in line 28, in each step we check whether $\context_S (i') \succ \context_T (j)$ for some $i' < i + k$ with
\[\BWT_S [i'] = \BWT_S [i] \neq \BWT_T [j]\]
and we output $\BWT_T [j]$ during the next pass.  The binary search in line 37 is symmetric.
\end{proof}

\noindent It follows from Lemma~\ref{lem:comparisons} that we can charge all the context comparisons to runs in $\BWT_{S, T}$ such that
\begin{itemize}
\item each run in $\BWT_{S, T}$ has $O (\log (m + n))$ context comparisons charged to it;
\item if a comparison between two characters' contexts is charged to a run in $\BWT_{S, T}$ then one of the characters is in that run and the other is not.
\end{itemize}
The number of characters we extract to compare the context of a character in a run $\BWT_{S, T} [a..b]$ to the context of a character in another run is at most
\[1 + \max \left( \rule{0ex}{2ex} \LCP_{S, T} [a], \LCP_{S, T} [b + 1] \right)\]
(unless $b = m + n$, in which case the number is just $\LCP_{S, T} [a]$), where $\LCP_{S, T} [1] = 0$ and, for $1 < h \leq m + n$, $\LCP_{S, T} [h]$ is the length of the longest common prefix of the contexts of the $h$th and $(h - 1)$st characters in $\BWT_{S, T}$.  This takes
\[O \left( \log r + \max \left( \rule{0ex}{2ex} \LCP_{S, T} [a], \LCP_{S, T} [b + 1] \right) \right)\]
time.  Therefore, the time for the context comparisons charged to $\BWT_{S, T} [a..b]$ is
\[O \left( \rule{0ex}{3ex} \left( \log r + \max \left( \rule{0ex}{2ex} \LCP_{S, T} [a], \LCP_{S, T} [b + 1] \right) \right) \log (m + n) \right)\,.\]
Summing over the $r$ runs in $\BWT_{S, T}$, the total time for context comparisons is
\[O ((r \log r + L) \log (m + n))\,,\]
where $L$ is the sum of its irreducible LCP values.  Combined with our previous observations, this gives us a preliminary result:

\begin{theorem}
\label{thm:first}
Given the RLBWTs $\BWT_S [1..m]$ and $\BWT_T [1..n]$, we can merge them into the run-length compressed eBWT $\BWT_{S, T} [1..m + n]$ using $O (r)$ space and $O ((r \log r + L) \log (m + n))$ time, where $m$ and $n$ are the lengths of the uncompressed strings, $r$ is the number of runs in $\BWT_{S, T}$ and $L$ is the sum of its irreducible LCP values.
\end{theorem}

\section{Proof of Lemma~\ref{lem:psi}}
\label{sec:construction}

The $\log r$ in Theorem~\ref{thm:first}'s time bound comes from paying the $O (\log r)$ overhead in Lemma~\ref{lem:psi} at every step of each binary search.  In Section~\ref{sec:optimization} we will explain how to avoid that and pay the overhead only three times during a binary search, but first it helps to examine the proof of Lemma~\ref{lem:psi}.

If an RLBWT for a string of length $n$ has $r$ runs then its $\Psi$ function is a permutation on $\{1, \ldots, n\}$ with
\[|\{1\} \cup \{i\ :\ 1 \leq i \leq n, \Psi (i) \neq \Psi (i - 1) + 1\}|
= r\,.\]
Given the RLBWT, in $O (r \log r)$ total time we can first build the set
\[\{(1, \Psi (1))\} \cup \{(i, \Psi (i))\ :\ 1 \leq i \leq n, \Psi (i) \neq \Psi (i - 1) + 1\}\,;\]
then apply Lemma~\ref{lem:splitting} below to obtain a slightly larger superset $P'$ with useful properties; and finally build a move structure with which, given $i$ and $i$'s predecessor in $P'$, in constant time we can find $\Psi (i)$ and $\Psi (i)$'s predecessor in $P'$.  Given only $i$, we can find $i$'s predecessor in $P'$ in $O (\log r)$, then use the move structure to iterate $\psi$ and extract the context of the $i$th character in the RLBWT character by character in constant time per character; Lemma~\ref{lem:psi} follows.

We refer readers to Nishimoto and Tabei's~\cite{NT21}, Brown et al.'s~\cite{BGR22} and Bertram et al.'s~\cite{BFN24} papers and Brown's~\cite{Bro23} thesis for more details on building move structures.  For the sake of completeness, we include a proof of Lemma~\ref{lem:splitting} in the appendix, mostly following Brown et al.'s and Bertram et al.'s presentation but with a potential-function argument by Alhadi~\cite{Alh26}.

\begin{lemma}
\label{lem:splitting}
Let $\pi$ be a permutation on $\{1, \ldots, n\}$ and
\[P = \{1\} \cup \{i\ :\ 1 < i \leq n, \pi (i) \neq \pi (i - 1) + 1\}\,.\]
Given the set $\{(i, \pi (i))\ :\ i \in P\}$ and an integer $d \geq 2$, we can build a set $\{(i, \pi (i))\ :\ i \in P'\}$, where
\begin{itemize}
	\item $P \subseteq P' \subseteq \{1, \ldots, n\}$,
    \item if $a$ and $b$ are consecutive elements in $\{\pi (i)\ :\ i \in P'\} \cup \{n + 1\}$ then $|[a, b) \cap P'| < 2 d$,
    \item $|P'| \leq \frac{d |P|}{d - 1}$.
\end{itemize}
This takes $O (|P| \log |P|)$ time in general but $O \left( \frac{|P| \log |P|}{d} \right)$ time when $n$ is polynomial in $|P|$.
\end{lemma}

\section{Optimization}
\label{sec:optimization}

The $O (\log r)$ overhead in Lemma~\ref{lem:psi} comes from a predecessor query on a set of size $O (r)$, as described in Section~\ref{sec:construction}.  For our algorithm, that is multiplied by the $O (r)$ passes we make through the while loop and by the $O (\log (m + n))$ context comparisons we may make during each pass if we reach line 8, 13, 28 or 37 and perform a binary search in an RLBWT.  If we do not perform such a binary search during a pass then we make only a constant number of context comparisons in that pass and they contribute only $O (\log r) \subseteq O (\log (m + n))$ overhead, so we need not worry about this case.

To see how to speed up the binary searches in the RLBWTs, consider how we search for the smallest value $j'$ with $\context_T (j') \succ \context_S (i + k)$ in line 8, for example.  We repeatedly choose candidate values $j'$ in $[j, j + \ell)$ and, for each one, compare $\context_T (j')$ with $\context_S (i + k)$.  If we choose the candidate values $j'$ na{\"i}vely, then each context comparison starts with a predecessor query on $P_T'$, where $P_T'$ is the set we build with Lemma~\ref{lem:splitting} as part of building a move structure for $\Psi$ on $T$.  If we choose $j' \in P_T'$ or already knowing the predecessor of $j'$ in $P_T'$, however, then we do not need such a predecessor query and we do not pay the $O (\log n)$ overhead for that step of the binary search.

In an optimized binary search for $j'$, we first pay the $O (\log r)$ overhead once to be able to extract $\context_S (i + k)$.  We then pay $O (\log r)$ overheads twice more to find the successor of $j$ and the predecessor of $j + \ell - 1$ in $P_T'$.  We then use binary search to find the smallest value $p \in [j, j + \ell) \cap P_T'$ with $\context_T (p) \succ \context_S (i + k)$, if there is one.  Since $p \in [j, j + \ell) \cap P_T' \subseteq P_T'$, we do not pay the $O (\log n)$ overhead for these steps.  Whether $p$ exists or not, we can now focus our search on interval of $[j, j + \ell)$ in which all the elements have the same predecessor in $P_T'$, so we do not pay the $O (\log n)$ overhead for these steps either.  Binary searches on $\BWT_S$ can be sped up symmetrically.

Lemma~\ref{lem:comparisons} still holds and we can charge all the context comparisons as before.  Now, however, the time for the context comparisons charged to a run $\BWT_{S, T} [a..b]$ in $\BWT_{S, T}$ is only
\[O \left( \rule{0ex}{3ex} \log r + \left( 1 + \max \left( \rule{0ex}{2ex} \LCP_{S, T} [a], \LCP_{S, T} [b + 1] \right) \right) \log (m + n) \right)\,.\]
Summing over the $r$ runs in $\BWT_{S, T}$, the total time for context comparisons is
\[O \left( \rule{0ex}{2ex} r \log r + (r + L) \log (m + n) \right)
= O ((r + L) \log (m + n))\,.\]
This gives us our main result:

\begin{theorem}
\label{thm:main}
Given the RLBWTs $\BWT_S [1..m]$ and $\BWT_T [1..n]$, we can merge them into the run-length compressed eBWT $\BWT_{S, T} [1..m + n]$ using $O (r)$ space and $O ((r + L) \log (m + n))$ time, where $m$ and $n$ are the lengths of the uncompressed strings, $r$ is the number of runs in $\BWT_{S, T}$ and $L$ is the sum of its irreducible LCP values.
\end{theorem}

\noindent
With care, it is possible to combine Theorems~\ref{thm:simple} and~\ref{thm:main} and obtain the advantages of each --- using doubling search among run boundaries and achieving adaptivity with respect to $b$, $B$ and $L$ simultaneously (better than simply dovetailing) --- but we leave that as an exercise for the reader until we publish the full version of this paper.

\appendix

\section{Proof of Lemma~\ref{lem:splitting}}

\setcounterref{theorem}{lem:splitting}
\addtocounter{theorem}{-1}

\begin{lemma}
Let $\pi$ be a permutation on $\{1, \ldots, n\}$ and
\[P = \{1\} \cup \{i\ :\ 1 < i \leq n, \pi (i) \neq \pi (i - 1) + 1\}\,.\]
Given the set $\{(i, \pi (i))\ :\ i \in P\}$ and an integer $d \geq 2$, we can build a set $\{(i, \pi (i))\ :\ i \in P'\}$, where
\begin{itemize}
	\item $P \subseteq P' \subseteq \{1, \ldots, n\}$,
    \item if $a$ and $b$ are consecutive elements in $\{\pi (i)\ :\ i \in P'\} \cup \{n + 1\}$ then $|[a, b) \cap P'| < 2 d$,
    \item $|P'| \leq \frac{d |P|}{d - 1}$.
\end{itemize}
This takes $O (|P| \log |P|)$ time in general but $O \left( \frac{|P| \log |P|}{d} \right)$ time when $n$ is polynomial in $|P|$.
\end{lemma}

\begin{proof}
Let $Q = \{\pi (i)\ :\ i \in P\} \cup \{n + 1\}$.  We build AVL trees $T_P$ and $T_Q$ storing the elements of $P$ and $Q$, respectively.  For each $i \in P$ we store $\pi (i)$ as satellite data with $i$ in $T_P$, and we store $i$ as satellite data with $\pi (i)$ in $T_Q$.  For each pair of consecutive elements $a$ and $b$ in $Q$, if $|[a, b) \cap P| \geq 2 d$ then we store the pair $(a, b)$ in a list $L$.  The bottleneck is sorting $P$ and $Q$, which takes $O (|P| \log |P|)$ time in general but $O (|P|)$ time when $n$ is polynomial in $|P|$.

We work in rounds and maintain the relationships between $T_P$, $T_Q$ and $L$.  In each round we check if $L$ is empty and, if so, we return each element $i$ in $T_P$ together with its satellite data $\pi (i)$ and then stop; if not, we remove 1 element from $L$ and then insert 1 element into $T_P$, 1 element into $T_Q$ and up to 2 elements into $L$.

Let $P_j$ and $Q_j$ be the contents of $T_P$ and $T_Q$ after $j$ rounds, so $P_0 = P$ and $Q_0 = Q$.  Suppose we remove $(s, e)$ from $L$ during the $(j + 1)$st round, meaning $[s, e) \cap P_j \geq 2 d$.  We use $T_P$ to find the $(d + 1)$st smallest element $p$ in $[s, e) \cap P_j$.  We use $T_Q$ to find $\pi^{-1} (p)$ by finding the predecessor $q$ of $p$ in $Q_j$ and adding $p - q$ to $\pi^{-1} (q)$, which is stored as satellite data with $q$.  We insert $\pi^{-1} (p)$ into $T_P$ with $p$ as satellite data, and insert $p$ into $T_Q$ with $\pi^{-1} (p)$ as satellite data.  Therefore $Q_{j + 1} = \{\pi (i)\ :\ i \in P_{j + 1}\} \cup \{n + 1\}$.

To see why $\pi^{-1} (p) = \pi^{-1} (q) + p - q$ and suppose $\pi (x) = y$.  If $x \in P_j$ then $y \in Q_j$, so if $y \not \in Q_j$ then $x \not \in P_j$.  Therefore, if $y \not \in Q_j$ then
$\pi (x) = \pi (x - 1) + 1$ and so $y - 1 = \pi (x - 1)$; applying $\pi^{-1}$ to both sides we have $\pi^{-1} (y - 1) = x - 1$ and so $\pi^{-1} (y) = \pi^{-1} (y - 1) + 1$.  Since $q$ is $p$'s predecessor in $Q_j$, it follows that $\pi^{-1} (p) = \pi^{-1} (q) + p - q$.

We use $T_P$ and $T_Q$ to check first whether $|[p, e) \cap P_{j + 1}| \geq 2 d$; if so, we add $(p, e)$ to $L$.  We then use $T_P$ and $T_Q$ to find the predecessor $s'$ of $\pi^{-1} (p)$ in $Q_{j + 1}$ and next element $e'$ after $s'$ in $Q_{j + 1}$, and to check whether $|[s', e') \cap P_{j + 1}| = 2d$; if so, we add $(s', e')$ to $L$.  Notice that if $|[s', e') \cap P_{j + 1}| > 2d$ then $|[s', e') \cap P_j| \geq 2d$ so $(s', e')$ was already in $L$.  This completes the $(j + 1)$st round, which takes $O (\log |P_j|)$ total time.

To bound the number of rounds we use, let
\[f (k) = \sum \left\{ \max \left( \rule{0ex}{2ex} |[a, b) \cap P_k| - d, 0 \right)\ :\ \mbox{$a$ and $b$ are consecutive elements in $Q_k$} \right\}\]
and consider how we obtain $f (j + 1)$ from $f (j)$.  We subtract the term $\max \left( \rule{0ex}{2ex} |[s, e) \cap P_j| - d, 0 \right)$; add the terms $\max \left( \rule{0ex}{2ex} |[s, p) \cap P_j| - d, 0 \right)$ and $\max \left( \rule{0ex}{2ex} |[p, e) \cap P_j| - d, 0 \right)$; and finally increment the term $\max \left( \rule{0ex}{2ex} |[s', e') \cap P_j| - d, 0 \right)$, to make everything with respect to $P_{j + 1}$ instead of $P_j$.  Since
\[2 d \leq |[s, e) \cap P_j|
= |[s, p) \cap P_j| + |[p, e) \cap P_j|
= d + |[p, e) \cap P_j|\,,\]
we have
\begin{eqnarray*}
\max \left( \rule{0ex}{2ex} |[s, e) \cap P_j| - d, 0 \right)
& = & |[p, e) \cap P_j|\,,\\[1ex]
\max \left( \rule{0ex}{2ex} |[s, p) \cap P_j| - d, 0 \right)
& = & 0\,,\\[1ex]
\max \left( \rule{0ex}{2ex} |[p, e) \cap P_j| - d, 0 \right)
& = & |[p, e) \cap P_j| - d\,.
\end{eqnarray*}
It follows that $f (j + 1) \leq f (j) - (d - 1)$.

Since $f (0) \leq |P|$ and we stop if $f (j + 1) = 0$ and $|P_{j + 1}| \leq |P_j| + 1$, we use at most $\frac{|P|}{d - 1}$ rounds and return $P'$ with
\[|P'|
\leq |P| + \frac{|P|}{d - 1}
= \frac{d |P|}{d - 1}\,.\]
Since the $(j + 1)$st round takes $O (\log |P_j|) = O (\log |P|)$ time for all $j$, over all the rounds we use $O \left( \frac{|P| \log |P|}{d} \right)$ total time.
\end{proof}

\end{document}